%% file: qge-ddc-rom_submitted.tex
\documentclass[11pt]{amsart}
\usepackage{amssymb, epsfig,amssymb, latexsym}
\usepackage{amsfonts,psfrag,amsmath,bbm,color,multirow,url}

\usepackage{algorithm}
\usepackage{algorithmicx}

\usepackage{placeins}
\usepackage{amscd}
\usepackage{bm}
\usepackage{csquotes}

\input{notation}

\usepackage[sort]{cite}
\usepackage[font=small,labelfont=bf]{caption}
\usepackage[normalem]{ulem} 
\definecolor{rred}{rgb}{0.7,0,0.1}
\definecolor{greenrb}{rgb}{0.2,0.6,0.2}

\definecolor{Carolinablue}{RGB}{115, 196, 251}

\graphicspath{{./figures/}}

\usepackage{fullpage}

\usepackage[foot]{amsaddr}
\makeatletter
\renewcommand{\email}[2][]{%
  \ifx\emails\@empty\relax\else{\g@addto@macro\emails{,\space}}\fi%
  \@ifnotempty{#1}{\g@addto@macro\emails{\textrm{(#1)}\space}}%
  \g@addto@macro\emails{#2}%
}
\makeatother
\begin{document}

\title{Data-Driven Correction Reduced Order Models \\ 
	for the Quasi-Geostrophic Equations: \\ 
	A Numerical Investigation
	}

\author[Changhong Mou]{Changhong Mou}
\address[CM, HL,TI]{Department of Mathematics, Virginia Tech, Blacksburg, VA 24061}
\email{cmou@vt.edu}
\author[Honghu Liu]{Honghu Liu}
\email{hhliu@vt.edu}
\author[David R. Wells]{David R. Wells}
\address[DRW]{Department of Mathematics, University of North Carolina, Chapel Hill, NC 27516}
\email{drwells@email.unc.edu}
\author[Traian Iliescu]{Traian Iliescu}
\email{iliescu@vt.edu}

\maketitle

\begin{abstract}
This paper investigates the recently introduced data-driven correction reduced order model (DDC-ROM) in the numerical simulation of the quasi-geostrophic equations.
The DDC-ROM uses available data to model the correction term that is generally used to represent the missing information in low-dimensional ROMs.
Physical constraints are added to the DDC-ROM to create the constrained
data-driven correction reduced order model (CDDC-ROM) in order to further improve its accuracy and stability.
Finally, the DDC-ROM is tested on time intervals that are longer than the time interval over which it was  trained.
The numerical investigation shows that, for low-dimensional ROMs, both the DDC-ROM and CDDC-ROM perform better than the standard Galerkin ROM (G-ROM) and the CDDC-ROM provides the best results.
\end{abstract}

\section{Introduction}
     \label{sec:introduction}

Reduced order models (ROMs) for fluid dynamics have been abundantly investigated in recent decades as a way to reduce the computational cost of high resolution numerical schemes. The success of many ROM approaches has already been documented for various scientific and engineering applications, especially for flows that are governed by relatively few recurrent dominant spatial structures; see e.g.~\cite{ballarin2016fast,fick2018stabilized,gunzburger2017ensemble,hesthaven2015certified,HLB96,noack2011reduced,perotto2017higamod,quarteroni2015reduced,sapsis2009dynamically,galletti2004low} and references therein. 

In this article, the recently proposed data-driven correction ROM (DDC-ROM) and its variants \cite{xie2018data,mohebujjaman2019physically,koc2019commutation} are investigated in the numerical simulation of a quasi-geostrophic model of the double-gyre wind-driven ocean circulation. The DDC-ROMs fall into the category of hybrid projection/data-driven ROMs \cite{couplet2005calibrated,galletti2004low,noack2005need,lu2017data,hijazi2019data}. More specifically, in DDC-ROMs, the interactions among the resolved modes are the same as those in the standard Galerkin projection ROMs, while the interactions involving the unresolved modes are learned through a data-driven approach by fitting, e.g., a quadratic ansatz to the data that represents these missing interactions. In the following, we provide a brief derivation of the DDC-ROMs that builds on the standard projection ROMs, and refer to \cite{xie2018data,mohebujjaman2019physically} for more details.

To construct the standard projection ROM, we start with a general nonlinear system that has the following weak form\footnote{subject to possible further integration by parts for certain terms in $( \bff(\bu) \, , \bv )$.} in a suitable Hilbert space $\bX$:
\begin{eqnarray}
	\bigl(  \overset{\bullet}{\bu} \, , \bv \bigr) 
	= \bigl( \bff(\bu) \, , \bv \bigr) \, ,
	\qquad 
	\forall \, \bv \in \bX \, ,	
	\label{eqn:equation}
\end{eqnarray}
where $\bff$ is a general nonlinear function, $\bu \in \bX$ is the sought solution, and $(\cdot, \cdot)$ denotes the inner product on $\bX$.
Next, we use available data (snapshots) to construct orthonormal modes $\{ \bphi_1, \ldots, \bphi_R \}$, which represent the recurrent spatial structures, where $R$ is the rank of the snapshot matrix and  typically $R = \cO(10^{3})$ or even higher.
Then, we choose the dominant modes $\{ \bphi_1, \ldots, \bphi_r \}$, typically with $r= \cO(10)$, as ROM basis functions.
The $r$-dimensional {\it Galerkin ROM (G-ROM)} of \eqref{eqn:equation} is obtained by replacing $\bu$ with a Galerkin truncation $\bur = \sum_{j=1}^{r} a_j \, \bphi_j$ and restricting $\bv$ to the ROM subspace $\bXr := \text{span} \{ \bphi_1, \ldots, \bphi_r \}$:
\begin{eqnarray}
	\bigl( \overset{\bullet}{\bu_{r}} \, , \bphi_{i} \bigr) 
	= \bigl( \bff(\bu_{r}) \, , \bphi_{i} \bigr) \, ,
	\qquad 
	i = 1, \ldots, r.	
	\label{eqn:g-rom-general}
\end{eqnarray}
In an offline stage, we construct the ROM, and in an online stage, we repeatedly use the G-ROM~\eqref{eqn:g-rom-general} with various parameters (if~\eqref{eqn:equation} is parameter dependent) and/or longer time intervals. 

To construct the {\it data-driven correction reduced order model (DDC-ROM)}~\cite{xie2018data,mohebujjaman2019physically,koc2019commutation}, we use an alternative  approach:
We start with a new Galerkin truncation, $\bu_{R} = \sum_{j=1}^{R} a_j \, \bphi_j$.
We emphasize that, since $R = \cO(10^{3})$ is the rank of the snapshot matrix, the new Galerkin truncation includes all the information in the available data (snapshots). 
Next, we replace $\bu$ with $\bu_{R}$ in~\eqref{eqn:equation} and project the resulting PDE onto $\bXr$:
\begin{eqnarray}
	\bigl( \overset{\bullet}{\bu_{R}} \, , \bphi_{i} \bigr) 
	= \bigl( \bff(\bu_{R}) \, , \bphi_{i} \bigr) \, ,
	\qquad 
	i = 1, \ldots, r.	
	\label{eqn:ddc-rom-1}
\end{eqnarray}
Since the ROM modes are orthonormal, 
$\bigl( \overset{\bullet}{\bu_{R}} \, , \bphi_{i} \bigr) 
= \bigl( \overset{\bullet}{\bu_{r}} \, , \bphi_{i} \bigr), \, 
i = 1, \ldots, r$.
Thus,~\eqref{eqn:ddc-rom-1} becomes
\begin{eqnarray}
	\bigl( \overset{\bullet}{\bu_{r}} \, , \bphi_{i} \bigr) 
	= \bigl( \bff(\bu_{R}) \, , \bphi_{i} \bigr) \, ,
	\qquad 
	i = 1, \ldots, r,	
	\label{eqn:ddc-rom-2}
\end{eqnarray}
which can be written as
\begin{eqnarray}
	\bigl( \overset{\bullet}{\bu_{r}} \, , \bphi_{i} \bigr) 
	= \bigl( \bff(\bu_{r}) \, , \bphi_{i} \bigr) 
		+
		\bigl[ 
			\bigl( \bff(\bu_{R}) \, , \bphi_{i} \bigr)
			-
			\bigl( \bff(\bu_{r}) \, , \bphi_{i} \bigr)
		\bigr] ,
	\qquad 
	i = 1, \ldots, r.
	\label{eqn:ddc-rom-3}
\end{eqnarray}
The last term on the right-hand side of~\eqref{eqn:ddc-rom-3} is a {\it Correction term}
\begin{eqnarray}
	\boxed{
	\text{Correction} 
	= \bigl[ 
			\bigl( \bff(\bu_{R}) \, , \bphi_{i} \bigr)
			-
			\bigl( \bff(\bu_{r}) \, , \bphi_{i} \bigr)
		\bigr] .
	}
	\label{eqn:correction}
\end{eqnarray}
Thus,~\eqref{eqn:ddc-rom-3} can be written as G-ROM + Correction.

We emphasize that~\eqref{eqn:ddc-rom-3} is expected to be more accurate than the G-ROM, since the former is constructed from $\bu_{R}$, whereas the latter is constructed from $\bu_{r}$, where $r = \cO(10) \ll R = \cO(10^{3})$.
Note that \eqref{eqn:ddc-rom-3} is not yet a closed system in $\bur$, since the Correction term involves $\bu_{R}$, which lives in a higher-dimensional space than $\bXr$. Thus, to obtain from \eqref{eqn:ddc-rom-3} an efficient $r$-dimensional ROM, we make the ansatz
\begin{eqnarray}
	\text{Correction} 
	= \bigl[ 
			\bigl( \bff(\bu_{R}) \, , \bphi_{i} \bigr)
			-
			\bigl( \bff(\bu_{r}) \, , \bphi_{i} \bigr)
		\bigr] 
		\approx
		\bigl( {\bf g}(\bu_{r}) \, , \bphi_{i} \bigr) \, ,
	\label{eqn:ansatz}
\end{eqnarray}
where ${\bf g}$ is a generic function (e.g., polynomial) whose coefficients/parameters still need to be determined.
Once ${\bf g}$ is determined, the ROM~\eqref{eqn:ddc-rom-3} with the Correction term replaced by ${\bf g}$ yields the {\it data-driven correction ROM (DDC-ROM)}:
\begin{eqnarray}
	\boxed{
	\bigl( \overset{\bullet}{\bu_{r}} \, , \bphi_{i} \bigr) 
	= \bigl( \bff(\bu_{r}) \, , \bphi_{i} \bigr) 
	+ \bigl( {\bf g}(\bu_{r}) \, , \bphi_{i} \bigr) ,
	}
	\qquad 
	i = 1, \ldots, r.
	\label{eqn:ddc-rom-general}
\end{eqnarray}
To determine the coefficients/parameters of the function ${\bf g}$ used in~\eqref{eqn:ddc-rom-general}, we use {\it data-driven modeling}~\cite{brunton2019data,loiseau2018constrained,peherstorfer2016data}, i.e., we solve the following {\it least squares problem}:
\begin{eqnarray}
	\boxed{
	\min_{ {\bf g}\text{ parameters}} \ \sum_{j=1}^{M} \left\| \text{Correction}(t_{j}) - \bigl( {\bf g}(\bu_{r}(t_{j})) \, , \bphi_{i} \bigr) \right\|^2 \, .
	}
	\label{eqn:least-squares}
\end{eqnarray}

The numerical investigations in~\cite{xie2018data,mohebujjaman2019physically,koc2019commutation} show that the DDC-ROM~\eqref{eqn:ddc-rom-general} is significantly more accurate than the standard G-ROM in the numerical simulation of two test problems:
(i) the 1D Burgers equation with a small diffusion coefficient $\nu=10^{-3}$; and 
(ii) a 2D flow past a circular cylinder at Reynolds numbers $Re=100, Re=500$, and $Re=1000$. 

\vspace*{0.3cm}

The main goal of this paper is to investigate the new DDC-ROM~\eqref{eqn:ddc-rom-general} in the numerical simulation of the {\it quasi-geostrophic equations (QGE)}, which represent a significantly more difficult test case than the Burgers equation and the 2D flow past a circular cylinder considered in \cite{xie2018data,mohebujjaman2019physically}. 
Indeed, for the 2D  flow past a circular cylinder with the Reynolds number $Re=1000$, 4the projection of the velocity field onto the first $8$ POD modes captures more than $99\%$ of the kinetic energy. 
In contrast, for the QGE in the parameter regime investigated below, a much broader range of spatial scales are actively involved in the time evolution of the turbulent fluid field. Indeed, the total amount of kinetic energy captured by the leading POD modes increases much slower for the QGE investigated here: it requires $16$ POD modes to capture $90\%$ of the kinetic energy, $37$ modes for $95\%$ of the kinetic energy, and $49$ modes for $96\%$ of the kinetic energy.

Furthermore,  given the challenges posed by the QGE, we investigate two improvements to the DDC-ROM~\eqref{eqn:ddc-rom-general}:
First, we study the role of adding {\it physical constraints} to the DDC-ROM~\cite{mohebujjaman2019physically}, in which the model for the Correction term in~\eqref{eqn:ansatz} satisfies the same type of physical constraints as those satisfied by the underlying equations; see \eqref{eqn:least-squares-qge-constrained}.
We also investigate whether modeling the {\it commutation error}, i.e., the error that appears as a result of interchanging spatial differentiation and ROM spatial filtering (e.g., projection)~\cite{koc2019commutation}, improves the DDC-ROM accuracy. Finally, we study the DDC-ROM when it is trained on a time interval that is shorter than the time interval over which it is tested.

\subsection{Connections to Previous Work} 
The DDC-ROM belongs to the class of {\it ROM closure models}, which model the effect of the truncated ROM modes (i.e., $\{ \bphi_{r+1}, \ldots, \bphi_{R} \}$) on the resolved ROM modes, (i.e., $\{ \bphi_{1}, \ldots, \bphi_{r} \}$).
ROM closure models were first proposed in the pioneering work of Lumley and his collaborators~\cite{HLB96} and are currently witnessing a dynamic development in several new directions, e.g., ROM spatial filtering, large eddy simulation (LES), and variational multiscale (VMS)~\cite{azaiez2017streamline,baiges2015reduced,bergmann2009enablers,wang2012proper,rebollo2017certified}, Mori-Zwanzig (MZ) formalism~\cite{parish2019adjoint}, nonlinear autoregression, moving averages with exogenous inputs (NARMAX)~\cite{chorin2015discrete,lu2017data}, multilevel approaches and empirical model reduction (EMR) \cite{kravtsov2005multilevel,majda2012physics,kondrashov2015data}, data-adaptive harmonic decomposition and multilayer Stuart-Landau models (DAH-MSLM) \cite{chekroun2017data,kondrashov2018multiscale}, and the parameterizing manifold (PM) approach rooted in the approximation theory of local invariant manifolds~\cite{CLW15_vol2,CLM19_closure}, to name just a few.
Probably the most dynamic development has been in using available data and machine learning techniques to develop ROM closure models~\cite{hijazi2019data,maulik2019time,pagani2019statistical,san2018neural,san2018extreme,xie2018data2,wan2018data}.

The DDC-ROM is a {\it hybrid projection/data-driven ROM}, in which the standard Galerkin method is used to model the terms involving only the resolved modes and available data is used to model only the ROM closure term.
This parsimonious/minimalistic data-driven approach is made possible by using ROM spatial filtering (i.e., ROM projection) and an LES/VMS framework to isolate the ROM closure term, which is then approximated by using data.
The DDC-ROM minimalistic data-driven framework is similar in spirit to the NARMAX \cite{chorin2015discrete,lu2017data} and PM \cite{CLW15_vol2,CLM19_closure} ROM closure models, although they differ in the way the closure terms are handled.  The DDC-ROM centers around ROM spatial filtering, whereas in the NARMAX approach the closure terms are modeled using nonlinear autoregression moving average with the resolved modes as exogenous inputs, and the PM approach parameterizes explicitly the unresolved modes in terms of the resolved modes.

\medskip

The rest of the paper is organized as follows:
In Section~\ref{sec:rom}, we present briefly the QGE and the construction of the corresponding DDC-ROM.
In Section~\ref{sec:numerical-experiments}, we assess the performance of the DDC-ROM using two metrics: the time-averaged streamfunction and the kinetic energy.
Finally, in Section~\ref{sec:conclusions}, we summarize our findings and outline future research directions.

\section{Data-Driven Correction ROM (DDC-ROM)}
    \label{sec:rom}

In this section, we present the construction of the DDC-ROM for the QGE.

\subsection{Quasi-geostrophic Equations (QGE)}
	\label{sec:qge}
	
In what follows, we use the {\it quasi-geostrophic equations (QGE)} as a mathematical model:
\begin{align}
\frac{\partial \omega}{\partial t}+J(\omega,\psi)-Ro^{-1}\frac{\partial \psi}{\partial x} &= Re^{-1} \Delta \omega+Ro^{-1}F,\label{eq:qge1}\\
\omega &=-\Delta \psi, \label{eq:qge2}
\end{align}
where $\omega$ is the vorticity, $\psi$ is the streamfunction, $Re$ is the Reynolds number, and $Ro$ is the Rossby number.  As a test problem for numerical investigation, we consider the QGE \eqref{eq:qge1}--\eqref{eq:qge2} with a symmetric double-gyre wind forcing given by 
\begin{align}
	F = \sin(\pi(y-1)) .
	\label{eq:qge:forcing}
\end{align}

The single-layer QGE~\eqref{eq:qge1}--\eqref{eq:qge2} (also known as a barotropic vorticity equation (BVE)), are a popular mathematical model for forced-dissipative large scale ocean circulation.
The idealized double-gyre wind forcing setting has been often used to understand the wind-driven circulation, e.g., the role of mesoscale eddies and their effect on the mean circulation.
ROMs for the QGE~\eqref{eq:qge1}--\eqref{eq:qge2} have been used, e.g., in~\cite{crommelin2004strategies,galan2008error,san2015stabilized,selten1995efficient,strazzullo2018model}.

The spatial domain of the QGE is $\Omega = [0,1]\times [0,2]$ and the time domain is $[0,80]$. 
We assume that $\psi$ and $\omega$ satisfy homogeneous Dirichlet  boundary conditions:
\begin{align} \label{eq:qge:bdry_cond}
\psi(t, x,y) =0,\qquad \omega(t, x,y)=0 \qquad \text{for}\quad (x,y)\in\partial \Omega \; \text{ and } \;  t \ge 0.
\end{align}

The QGE~\eqref{eq:qge1}--\eqref{eq:qge2} can be cast in the general form of the nonlinear equation~\eqref{eqn:equation} by choosing
\begin{eqnarray}
	\left( f(\omega) , v \right)
	=  
	- \left( J(\omega,\psi) , v \right)
	+ Ro^{-1} \left( \frac{\partial \psi}{\partial x} , v \right)
	- Re^{-1} \left( \nabla \omega , \nabla v \right)
	+ Ro^{-1} \left( F , v \right) .
	\label{eqn:qge-f}
\end{eqnarray}
We refrain from giving the precise formulation of the functional space here since this is tangential to the numerical study carried out below. The interested readers can consult for instance \cite[Chapter 11]{Gun89}, for the case of Navier-Stokes equations in the streamfunction-vorticity formulation. We will make precise how each of the terms in \eqref{eqn:qge-f} is computed numerically once a set of POD basis functions for the vorticity is computed based on the direct numerical simulation (DNS) data obtained from a spectral code; see Sections \ref{sec:g-rom} and \ref{Sect_snapshot_generation}.

\subsection{Standard Galerkin ROM (G-ROM)}
    \label{sec:g-rom}

In our investigation, the ROM basis is obtained by using the proper orthogonal decomposition (POD)\cite{HLB96,noack2011reduced}. We note, however, that other bases could be used for this purpose as well; examples include the dynamic mode decomposition \cite{schmid2010dynamic}, the principal interaction patterns \cite{hasselmann1988pips,kwasniok1996reduction,kwasniok1997optimal}, and the HIGAMod~\cite{perotto2017higamod}. See also \cite{taira2017modal,crommelin2004strategies,tu2013dynamic} for recent surveys and relationships/comparisons between different modal decomposition approaches.

We focus here mainly on the functional form of the G-ROM and defer details about the POD basis construction to Section~\ref{sec:numerical-experiments}. To this end, given an $r$-dimensional ROM subspace $X^{r}$ spanned by the first $r$ POD basis functions for the vorticity $\omega$,  
\begin{equation} \label{eq:pod_vorticity}
X^{r} := \text{span} \{ \varphi_1, \ldots, \varphi_r \},
\end{equation} 
the $r$-dimensional G-ROM takes the form of \eqref{eqn:g-rom-general} with $f$ therein given by \eqref{eqn:qge-f}. Recall that the streamfunction $\psi$ in \eqref{eqn:qge-f} is related to the vorticity $\omega$ through the Poisson equation \eqref{eq:qge2} subject to homogeneous Dirichlet boundary conditions. 

To further reduce the G-ROM to an explicit ODE system, one option would be to replace $\psi$ in \eqref{eqn:qge-f} by $-\Delta^{-1}\omega$,  with $\Delta^{-1}$ being the inverse of the Laplacian of $\omega$ subject to the aforementioned boundary conditions. But since one important metric we adopt to assess the performance of the ROMs concerns the time average of $\psi$, we decide to keep $\psi$ explicit in the ROM formulation, although either way would lead to the same $r$-dimensional ODE system. For this reason, we also introduce  a reduced set of $r$ basis functions for $\psi$, which are  subordinate to the above POD basis functions for $\omega$ in \eqref{eq:pod_vorticity} via $\phi_i(x,y)  = - \Delta^{-1} \varphi_i(x,y)$, i.e., they solve the following Poisson equation:
\begin{align}
 - \Delta \phi_i(x,y)  = \varphi_i(x,y), \quad \text{ subject to } \quad  \phi_i(x,y) =0, \quad   \text{for}\quad (x,y)\in\partial \Omega, \qquad i=1,2,\cdots,r. 
\label{eqn:basis-streamfunction}
\end{align}
Note that while the POD basis $\{\varphi_i\}$ for the vorticity $\omega$ is an orthonormal basis under $L^2$-inner product, the basis $\{\phi_i\}$ for the streamfunction $\psi$ is not orthogonal.  
Given the G-ROM approximation $\omega_r = \sum_{i=1}^r a_i(t) \varphi_i(x,y)$ of $\omega$, the corresponding $\psi$ is approximated by $\psi_r = \sum_{i=1}^r a_i(t) \phi_i(x,y)$, which results from the ansatz $\psi_r = -\Delta^{-1}\omega_r$ and the above definition of the basis function $\phi_i$. 

With the above notations, the $r$-dimensional G-ROM for the problem \eqref{eq:qge1}--\eqref{eq:qge:bdry_cond} is given by:
\begin{align}
\left(\frac{\partial \omega_r}{\partial t},\varphi_i \right)+\left( J(\omega_r,\psi_r),\varphi_i\right)-Ro^{-1}\left(\frac{\partial \psi_r}{\partial x} ,\varphi_i\right)+Re^{-1} \left( \nabla\omega_r,\nabla\varphi_i\right)=Ro^{-1}\left(F,\varphi_i\right) ,
\label{eq:rom:qge}
\end{align}
where $(\cdot,\cdot)$ denotes the $L^2$ inner product over the spatial domain, and $i = 1,\cdots, r$.
Plugging the vorticity and streamfunction ROM approximations in~\eqref{eq:rom:qge}, yields the {\it Galerkin ROM (G-ROM)}:  
\begin{eqnarray}
	\dot{\ba}
	= 
	\bb
	+ A \, \ba
	+ \ba^{\top} \, B \, \ba \, ,
	\label{eqn:g-rom-qge}
\end{eqnarray}
where $\ba\stackrel{def}{=} \bigl( a_{j}(t) \bigr)_{j=1}^{r}$ is the vector of time-varying ROM coefficients.
The G-ROM~\eqref{eqn:g-rom-qge} can be written componentwise as follows: 
for $i=1,2,\cdots,r$
\begin{align}
\dot{a}_i(t) = b_i +\sum_{m=1}^r A_{im}a_m(t) +\sum_{m=1}^r\sum_{n=1}^rB_{imn} a_m(t)a_n(t) , 
\end{align}
where 
$
b_i = Ro^{-1} (F,\varphi_i), \, 
A_{im } = Ro^{-1} \left(\frac{\partial \phi_m}{\partial x},\varphi_i\right)-Re^{-1}\left(\nabla\varphi_m,\nabla\varphi_i\right), \, 
B_{imn}= -\left(J(\varphi_m, \phi_n),\varphi_i\right) .
$

\subsection{DDC-ROM}
	\label{sec:ddc-rom-construction}
	
To construct the DDC-ROM~\eqref{eqn:ddc-rom-general} for the QGE, we adapt the general presentation in Section~\ref{sec:introduction} to the QGE setting.

First, we note that the Correction term~\eqref{eqn:ansatz} takes the following form for the QGE: 
\begin{eqnarray}
	\text{Correction} 
	&=& \biggl( f(\omega_{R}) - f(\omega_{r}) \, , \varphi_{i} \biggr)
	\nonumber \\[0.3cm]
	&=&
	\biggl( J(\omega_R , \psi_R) - J(\omega_r , \psi_r) , \varphi_i \biggr)
	\nonumber \\[0.3cm]
	&=&
	\left( \frac{\partial\omega_R }{\partial x}\frac{\partial\psi_R }{\partial y}- \frac{\partial\psi_R }{\partial x}\frac{\partial\omega_R }{\partial y} , \varphi_i\right) 
	- 
	\left( \frac{\partial\omega_r }{\partial x}\frac{\partial\psi_r }{\partial y}- \frac{\partial\psi_r }{\partial x}\frac{\partial\omega_r }{\partial y},\varphi_i \right) ,
	\label{eqn:correction-qge}
\end{eqnarray}
where $\omega_R({\bf x},t) = \sum_{i=1}^{R} a_i(t) \varphi_i({\bf x})$ and $\psi_R({\bf x},t) = \sum_{i=1}^{R} a_i(t) \phi_i({\bf x})$ are the $R$-dimensional ROM approximations of the vorticity and streamfunction in $X^{R}$, respectively, and ${\bf x} = (x,y)$. 
For clarity of presentation, in~\eqref{eqn:correction-qge} we assume that the differentiation and the ROM projection commute (see, however,~\cite{koc2019commutation} for a detailed discussion of the commutation error).
Thus, the linear terms in the QGE do not appear in the Correction~\eqref{eqn:correction-qge}. 
We emphasize that the Correction~\eqref{eqn:correction-qge} is $R$-dimensional instead of $r$-dimensional, where $R \gg r$.
Thus, to include the Correction~\eqref{eqn:correction-qge} in the DDC-ROM, we first need to find an efficient, $r$-dimensional approximation of the Correction.
To this end, we make the following {\it linear} ansatz:
$\forall \, i = 1, \ldots, r,$
\begin{eqnarray}
	\text{Correction} 
	&=&
	\left( \frac{\partial\omega_R }{\partial x}\frac{\partial\psi_R }{\partial y}- \frac{\partial\psi_R }{\partial x}\frac{\partial\omega_R }{\partial y} , \varphi_i\right) 
	- 
	\left( \frac{\partial\omega_r }{\partial x}\frac{\partial\psi_r }{\partial y}- \frac{\partial\psi_r }{\partial x}\frac{\partial\omega_r }{\partial y},\varphi_i \right) 
	\nonumber \\[0.3cm]
	&\approx& \bigl( g(\omega_{r}) \, , \varphi_{i} \bigr) 
	\nonumber \\[0.3cm]
	&=& \bigl( \tilde{A} \, \ba \bigr)_{i} \ ,
	\label{eqn:ansatz-qge}
\end{eqnarray}
where the operator $\tA \in \R^{r \times r}$ needs to be determined and $\bigl( \tilde{A} \, \ba  \bigr)_{i}$ denotes the $i$-th component of the vector $\bigl( \tilde{A} \, \ba \bigr)$.
The ansatz~\eqref{eqn:ansatz-qge} is chosen to resemble the right-hand side of the G-ROM~\eqref{eqn:g-rom-qge}; we note, however, that other ansatzes are possible~\cite{xie2018data,mohebujjaman2019physically}. 

To compute the entries in the operator $\tA$ in~\eqref{eqn:ansatz-qge}, we use a data-driven approach.  
To this end, we adapt the least squares problem~\eqref{eqn:least-squares} to the QGE setting: 
\begin{eqnarray}
	\qquad \min_{\tA} \ \sum_{j=1}^{M} 
	\, \biggl\| 
		\biggl [
	\left( \frac{\partial\omega_R }{\partial x}\frac{\partial\psi_R }{\partial y}- \frac{\partial\psi_R }{\partial x}\frac{\partial\omega_R }{\partial y} , \varphi_i\right) 
	- 
	\left( \frac{\partial\omega_r }{\partial x}\frac{\partial\psi_r }{\partial y}- \frac{\partial\psi_r }{\partial x}\frac{\partial\omega_r }{\partial y},\varphi_i \right) 
		\biggr ]
	-  \tilde{A} \, \ba^{DNS}(t_{j})  
	\biggr\|^2 \, .
	\label{eqn:least-squares-qge}
\end{eqnarray}
In~\eqref{eqn:least-squares-qge}, $\ba^{DNS}(t_{j})$ is the vector of ROM coefficients obtained from the DNS data, i.e., from the snapshots, at time instances $t_j , \ j = 1, \ldots, M$, which are obtained by projecting the corresponding snapshots $\omega^{DNS}(t_j) = \sum_{k=1}^{R} a^{DNS}_k(t_j) \, \varphi_k$ onto the POD basis functions $\varphi_i$ and using the orthogonality of the POD basis functions: 
$\forall \, i = 1, \ldots, r, \ \forall \, j = 1, \ldots, M,$
\begin{equation}
	a^{DNS}_i(t_j)
	= \biggl( \omega^{DNS}(t_j) , \varphi_i \biggr) \, .
	\label{eqn:a-dns}
\end{equation}

The \textit{data-driven correction ROM (DDC-ROM)} has the following form for the QGE:
\begin{eqnarray}
	\dot{\ba} 
	= 
	\bb
	+ ( A + \tilde{A}) \ba 
	+ \ba^\top B  \ba \, ,
	\label{eqn:ddc-rom-qge}
\end{eqnarray}
where the operators $\bb, A$, and $B$ are the G-ROM operators in \eqref{eqn:g-rom-qge} and the operator $\tilde{A}$ is the solution of the least squares problem \eqref{eqn:least-squares-qge}.

\section{Numerical Experiments}
	\label{sec:numerical-experiments}

In this section, we investigate the DDC-ROM~\eqref{eqn:ddc-rom-qge} in the numerical simulation of the QGE.

%
%

\subsection{Computational Setting and Snapshot Generation}
	\label{Sect_snapshot_generation}

We investigate the QGE~\eqref{eq:qge1}--\eqref{eq:qge2} with the symmetric double-gyre wind forcing given in~\eqref{eq:qge:forcing} and homogeneous Dirichlet  boundary conditions for $\psi$ and $\omega$ given in~\eqref{eq:qge:bdry_cond}.
The parameters are set to be $Re=450$ and $Ro=0.0036$. 

For the DNS spatial discretization, we use a spectral method with a $257 \times 513$ spatial resolution. Since both the vorticity and streamfunction have homogeneous boundary conditions, we approximate both functions with a tensor product Sine expansion in \(x\) and \(y\).
For the DNS time discretization, we use an explicit Runge-Kutta method (Tanaka-Yamashita, an order $7$ method with an embedded order $6$ method for error control) and an error tolerance of $10^{-8}$ in time with adaptive time refinement and coarsening. We record the solution values every \(10^{-2}\) simulation
time units (starting at \(0\)) regardless of the current time step size so that
the snapshots used in the POD are equally spaced.
These spatial and temporal discretizations yield numerical results that are similar to the fine resolution numerical results obtained in~\cite{san2015stabilized,san2011approximate}. 
We follow~\cite{san2015stabilized,san2011approximate} and run the DNS between  $[0, 80]$.
The time evolution of the spatially averaged kinetic energy in Fig.~1 in~\cite{san2015stabilized} shows that the flow converges to a statistically steady state, after a short transient interval that ends around $t = 10$.
We emphasize that even in the statistically steady state regime, the flow displays a high degree of variability.
Thus, the numerical approximation of this statistically steady regime remains challenging for the low resolution ROMs that we investigate in this section.

\subsection{ROM Construction}
	\label{sec:rom-construction}

To generate the ROM basis (see Section~\ref{sec:rom}), we follow~\cite{san2015stabilized,san2011approximate} and collect $701$ equally spaced snapshots of the vorticity, $\omega$, in the time interval $[T_{min},T_{max}] = [10,80]$ (on which the statistically steady state regime is attained) at equidistant time intervals.
We also interpolate the DNS vorticity onto a uniform mesh with the resolution $257\times 513$ over the rectangle domain $\Omega = [0,1]\times [0,2]$, i.e., $h = \Delta x =\Delta y = 1/256$. We then use the $701$ snapshots, form the correlation matrix $C$ for vorticity, and obtain the POD basis functions $\varphi_i$'s from the eigenvectors of $C$. Recall that the element $C_{ij}$ of $C$ is simply the $L^2$ inner product of the $i$-th and the $j$-th snapshots, i.e., $C_{ij} = \int_{\Omega} \omega_i \omega_j \mathrm{d} x \mathrm{d} y$~\cite{san2015stabilized}. Throughout the article, the $L^2$ inner product over $\Omega$ is carried out by using the two-dimensional form of Simpson's $1/3$ rule. Once the POD basis functions for the vorticity are generated, we solve the Poisson equation~\eqref{eqn:basis-streamfunction} to construct the streamfunction basis functions.  For this purpose, we use a second order central difference (five-point stencil) spatial discretization of the Laplace operator.

\begin{figure}[htp]
\centering
\includegraphics[width=0.32\textwidth]{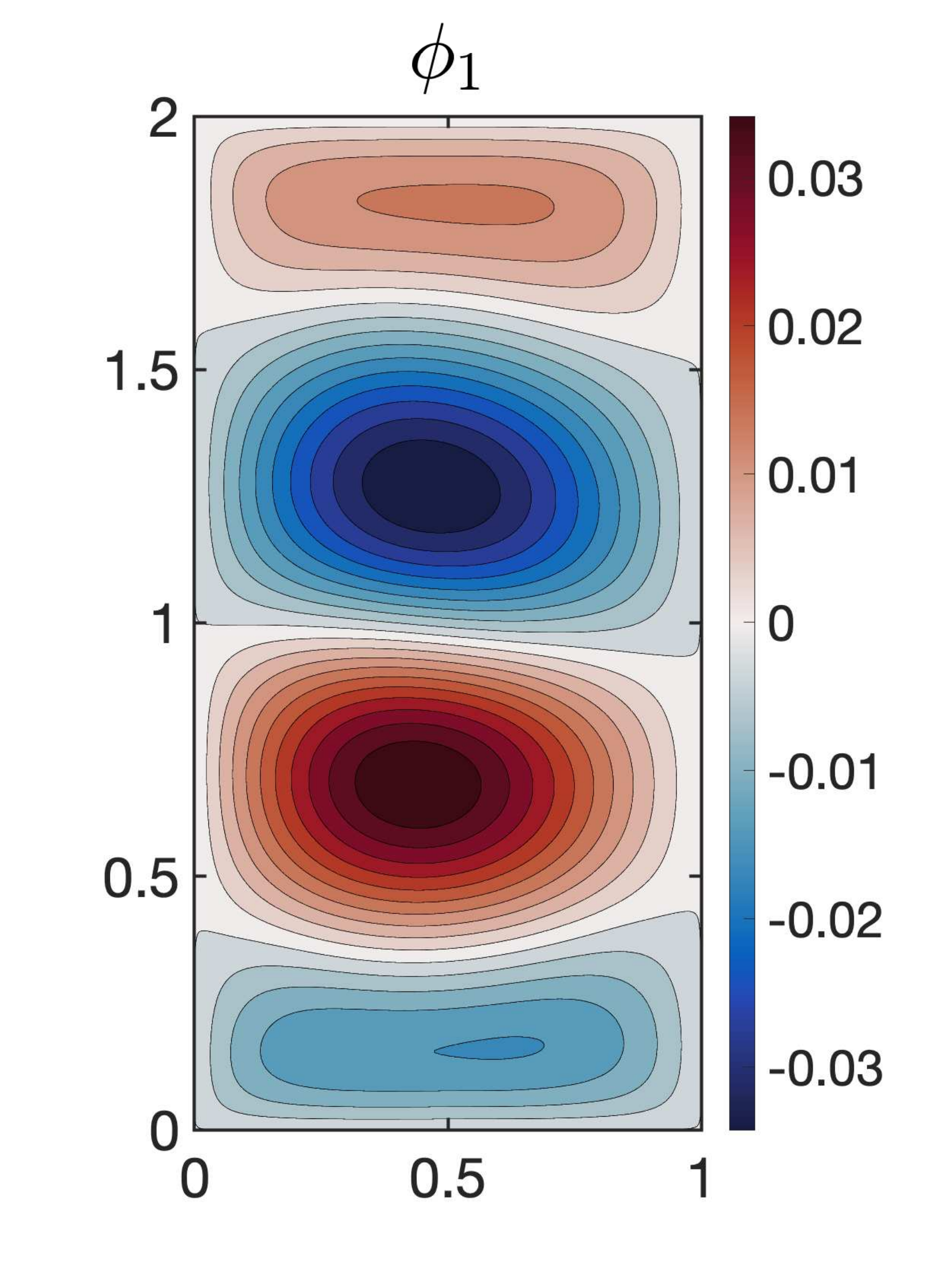}
\includegraphics[width=0.32\textwidth]{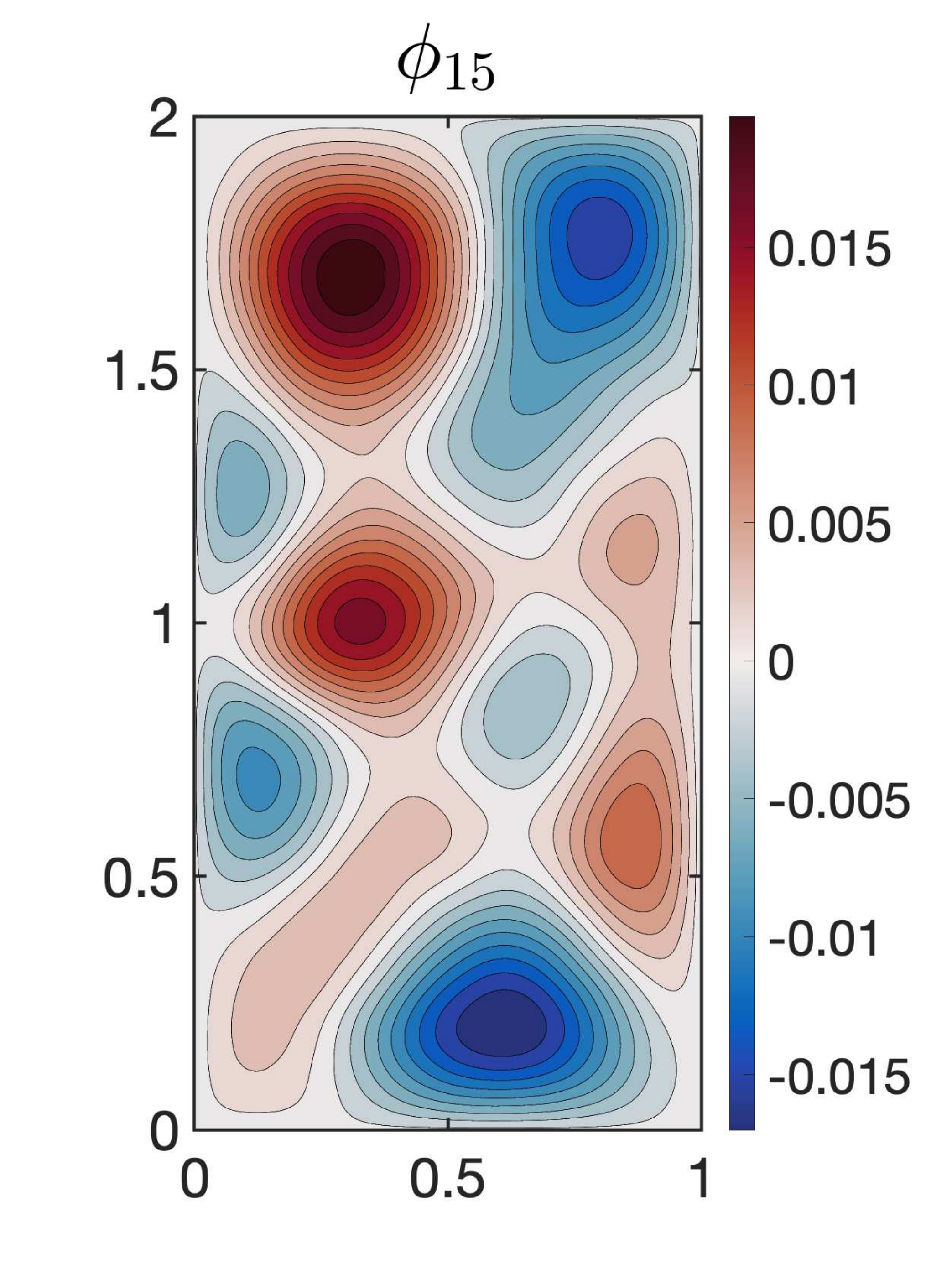}
\includegraphics[width=0.32\textwidth]{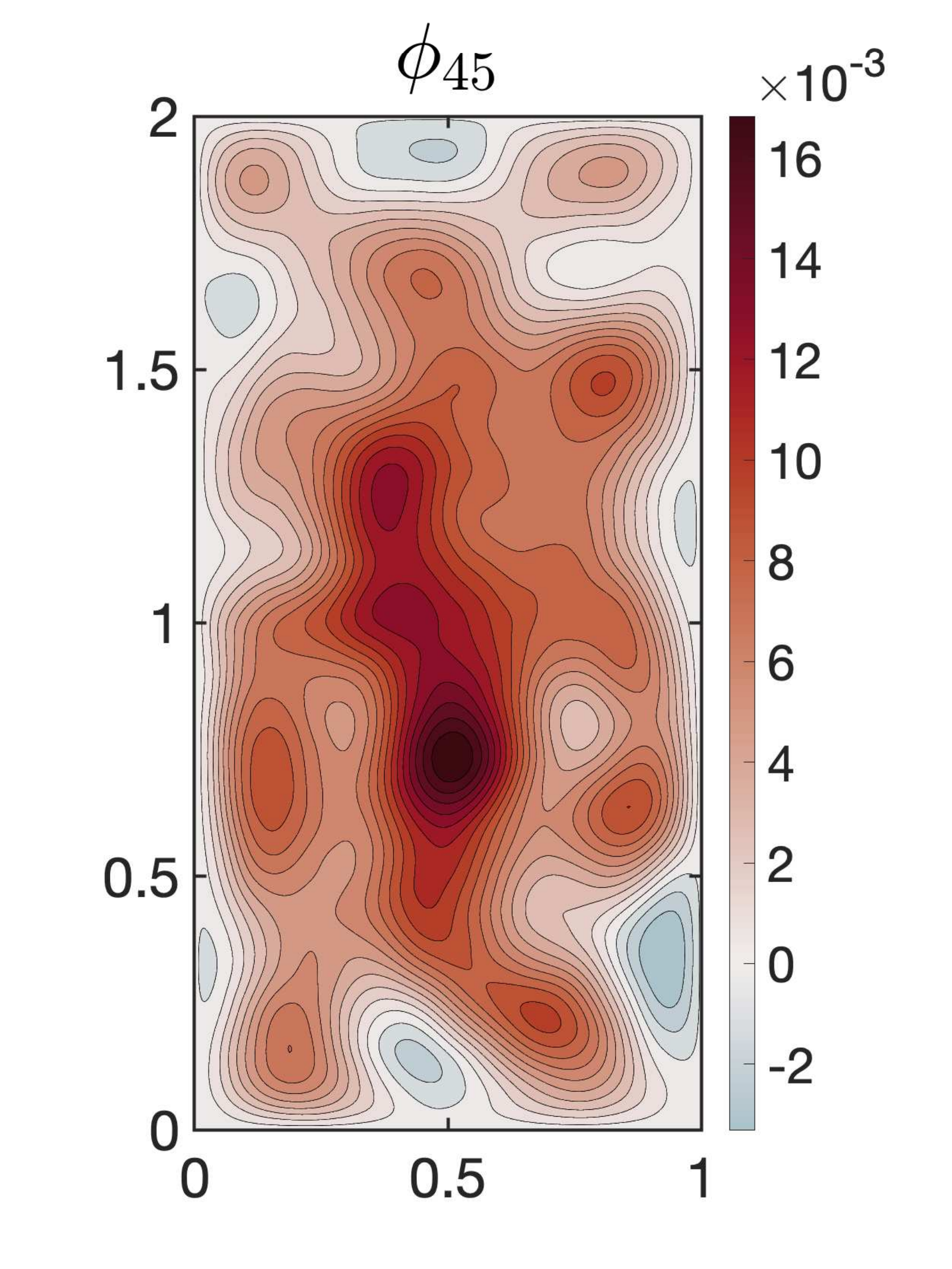}
\includegraphics[width=0.32\textwidth]{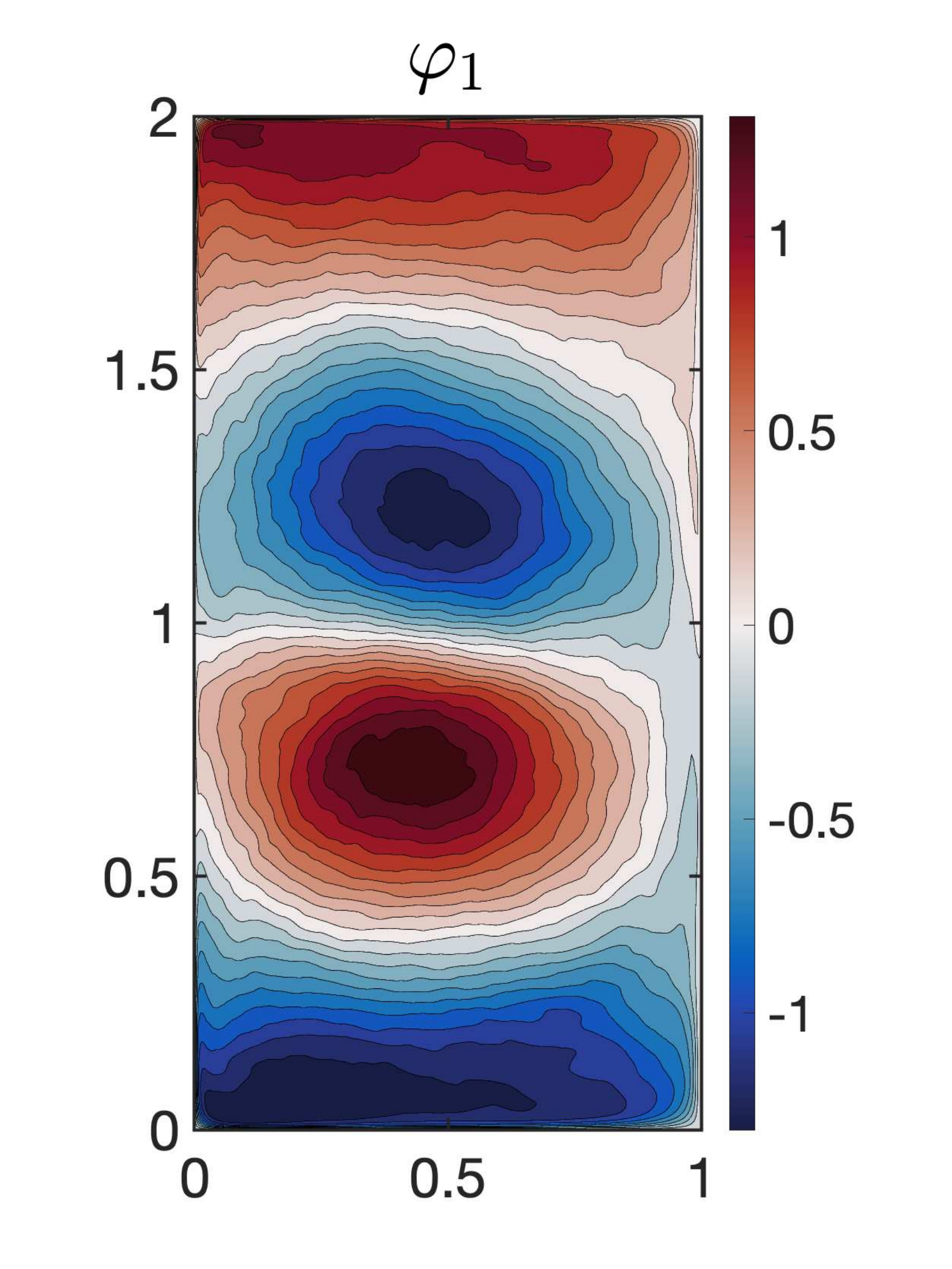}
\includegraphics[width=0.32\textwidth]{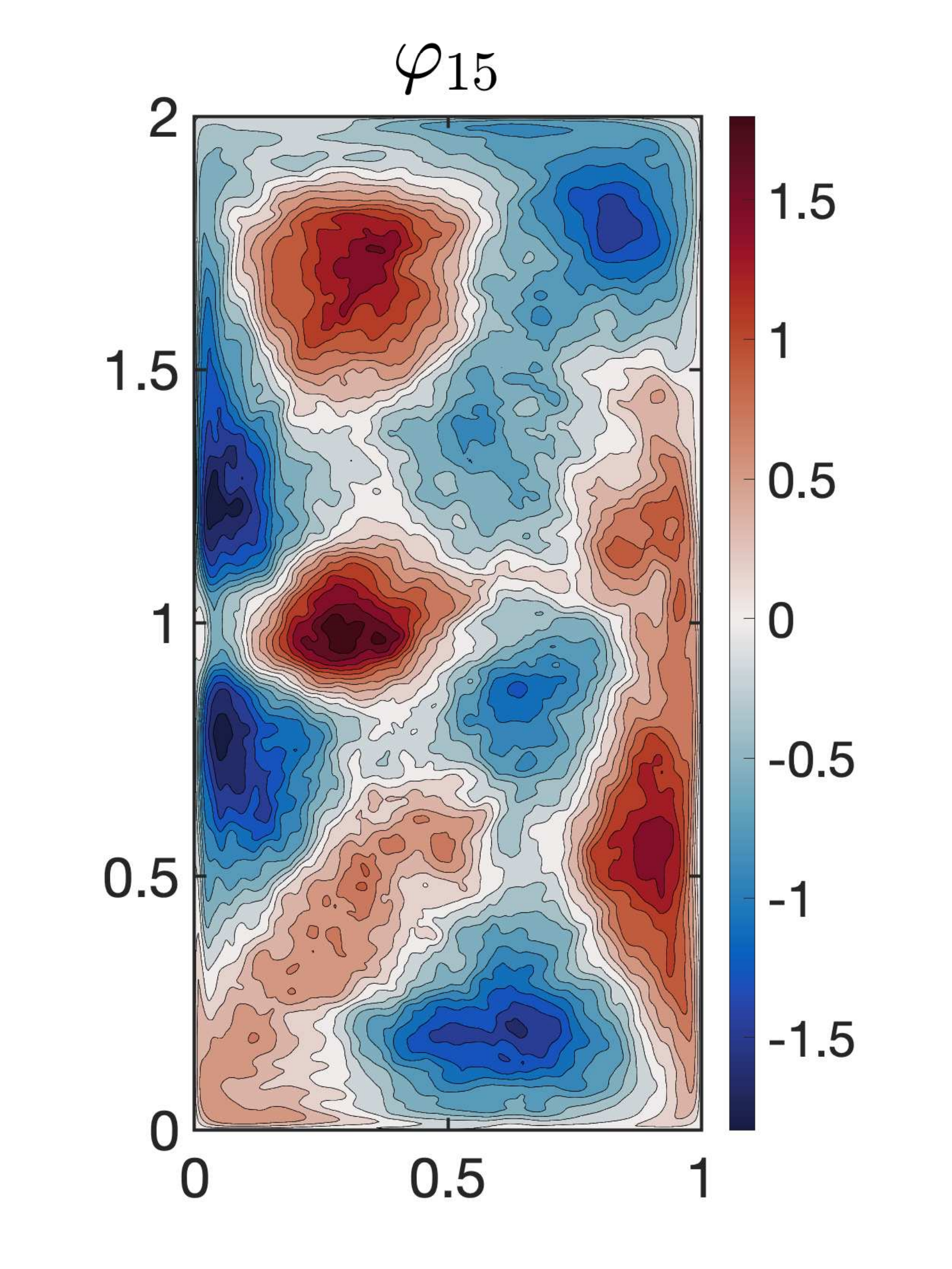}
\includegraphics[width=0.32\textwidth]{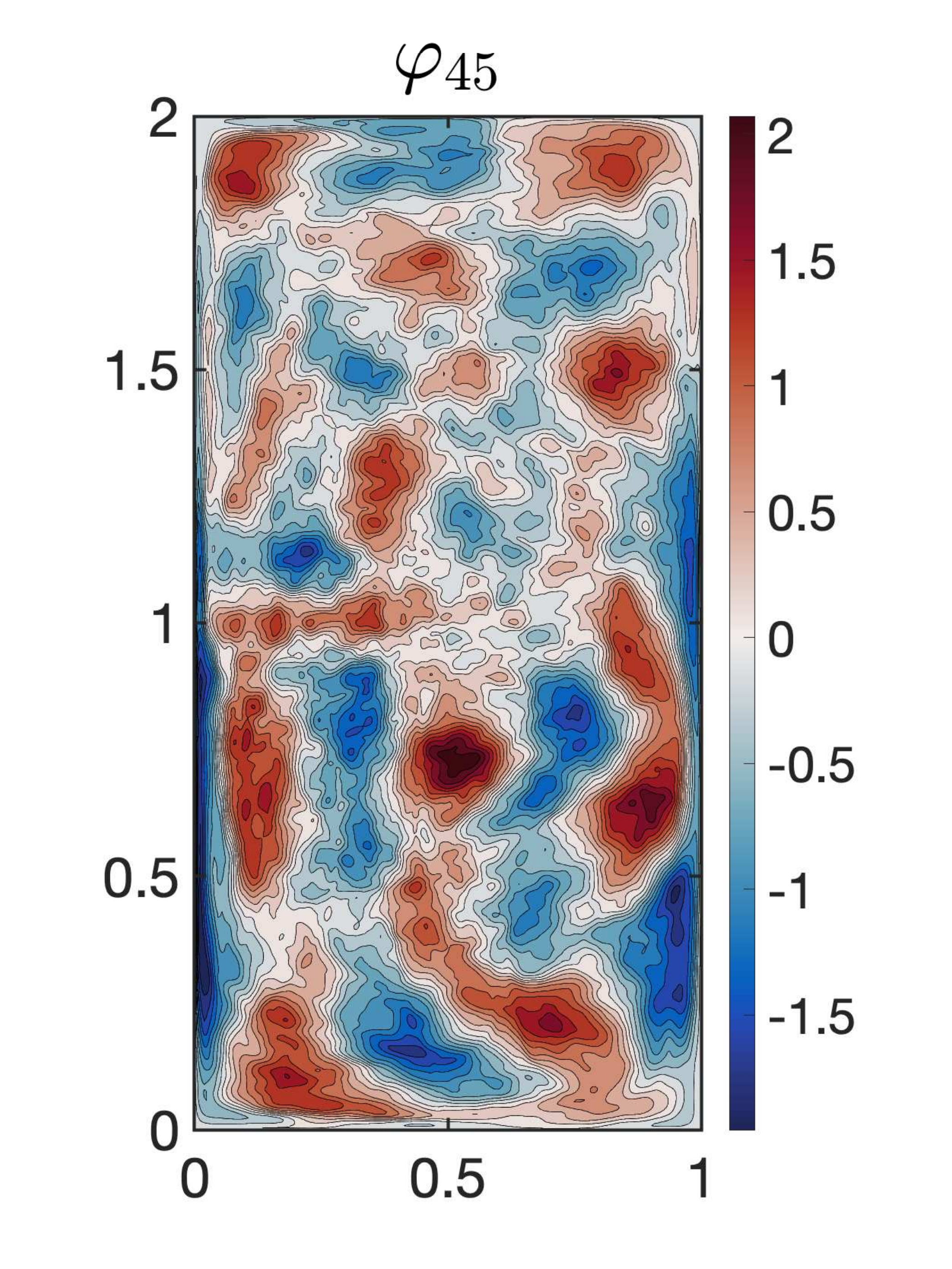}
\caption{
	Basis functions for the streamfunction (first row) and vorticity (second row). The vorticity basis functions $\varphi_i$'s are the POD modes computed based on the DNS snapshots for the vorticity, while each streamfunction basis function $\phi_i$ is related to $\varphi_i$ via $\phi_i = - \Delta^{-1} \varphi_i$; see Section~\ref{sec:g-rom}.
	\label{fig:pod-basis}
}
\end{figure}

In Fig.~\ref{fig:pod-basis}, we present the contour plots of selected streamfunction basis functions $\phi_i$'s and vorticity POD basis functions $\varphi_i$'s to give an idea about how the spatial scales are organized in these computed bases as the basis function index increases. Note that $\varphi_i$'s are much rougher than $\phi_i$'s, especially for the higher indices, whereas the roughness is smoothed out by the Laplacian when the $\phi_i$'s are computed according to~\eqref{eqn:basis-streamfunction}. One source of roughness could be the uniform discretization mesh adopted here ($257\times 513$) when interpolating the DNS data, since there are steep gradients in the vorticity field as time evolves, both near the western boundary and within the domain. However, as we will illustrate in the next section, such roughness does not significantly degrade the DDC-ROM accuracy.

To construct the DDC-ROM, we need to solve the least squares problem~\eqref{eqn:least-squares-qge}, which can be ill-conditioned, especially when the training data is relatively short with respect to the number of coefficients that need to be learnt. To tackle this ill-conditioning issue, we use the truncated singular value decomposition (SVD) (see Step 6 of Algorithm 1 in~\cite{xie2018data}) with a tolerance that yields the most accurate results.
Furthermore, to increase the computational efficiency of the DDC-ROM, we replace $( \omega_{R} , \psi_{R} )$ in~\eqref{eqn:least-squares-qge} with $( \omega_{m} , \psi_{m} )$, where $1 \leq m \leq R$ (for details, see Section 5.3 in~\cite{xie2018data}).
Our numerical experiments suggest that $m=3r$ achieves a good balance between numerical accuracy and computational efficiency for the considered QGE model.

In our numerical investigation, in addition to the DDC-ROM, we also consider the {\it physically-constrained} DDC-ROM (CDDC-ROM)~\cite{mohebujjaman2019physically}, which aims at improving the physical accuracy of the DDC-ROM.
To construct the CDDC-ROM, we add physical constraints that require that the data-driven CDDC-ROM operators satisfy the same type of physical laws as those satisfied by the QGE.
Specifically, we require that the CDDC-ROM's Correction term's linear component (i.e., the matrix $\tilde{A}$) should be dissipative.
To implement these physical constraints, in the data-driven modeling step, we replace the unconstrained least squares problem~\eqref{eqn:least-squares-qge} with a constrained least squares problem:
\begin{equation}
	\min_{
		\substack{\tA \in \R^{r \times r} \\[0.1cm] 
				\ba^{\top} \tA \, \ba \leq 0 
				}
		} \, 
	\ \sum_{j=1}^{M} 
	\, \biggl\| 
		\biggl [
	\left( \frac{\partial\omega_R }{\partial x}\frac{\partial\psi_R }{\partial y}- \frac{\partial\psi_R }{\partial x}\frac{\partial\omega_R }{\partial y} , \varphi_i\right) 
	- 
	\left( \frac{\partial\omega_r }{\partial x}\frac{\partial\psi_r }{\partial y}- \frac{\partial\psi_r }{\partial x}\frac{\partial\omega_r }{\partial y},\varphi_i \right) 
		\biggr ]
	-  \tilde{A} \, \ba^{DNS}(t_{j})  
	\biggr\|^2 \, .
	\label{eqn:least-squares-qge-constrained}
\end{equation}

Additionally, we also monitor the commutation error, which represents the effect of interchanging ROM spatial filtering and differentiation~\cite{koc2019commutation}. 
For this test problem, modeling the commutation error does not significantly change the DDC-ROM and CDDC-ROM results, suggesting that the commutation error does not play a significant role in the ROM construction.
Thus, for clarity of presentation, we do not include the commutation error in the DDC-ROM and CDDC-ROM results.

In the online stage, for all the ROMs we utilize the fourth order Runge-Kutta scheme (RK4) for the temporal discretization. 
To ensure the numerical stability of the time discretization, we choose a time step size $\Delta t =0.001$. 
We store ROM data every ten time steps to match the DNS sampling rate.
We use the DNS snapshot at $t=10$ to initialize the ROMs.

\subsection{Numerical Results}

In this section, we present numerical results for the G-ROM, DDC-ROM, and CDDC-ROM.
As benchmark for our numerical investigation, we use the DNS results.

\subsubsection{Kinetic energy}
	\label{sect_kinetic_energy}

In this section, we assess the performance of the ROMs using the DNS kinetic energy as a metric. As pointed out in Section~\ref{sec:introduction}, due to the involvement of a broad range of active spatial scales, it would be too demanding to require any ROMs to reproduce the statistics of the DNS kinetic energy or any other reasonable observables when the dimension is too low, at least within the POD basis framework adopted here. Thus, we confine ourselves instead to a much less ambitious goal of reproducing the range of oscillations presented in the DNS kinetic energy. The assessment at a more quantitative level will be carried out in the next section for another metric. 

Recall that the velocity field $(u(\bx,\cdot),v(\bx,\cdot))$ used in the computation of the kinetic energy $E(t) = \frac{1}{2}\int_\Omega \big( u^2(\bx,t) + v^2(\bx,t) \big) d\bx$ is related to the streamfunction according to $(u,v) = (\partial_y\psi,-\partial_x\psi)$. The first-order spatial derivatives are calculated using a $4$-th order accurate central difference scheme. The kinetic energy itself is computed using the two-dimensional form of Simpson's $1/3$ rule.

\begin{figure}[htp]
\centering
\includegraphics[width=0.99\textwidth,height=0.25\textwidth]{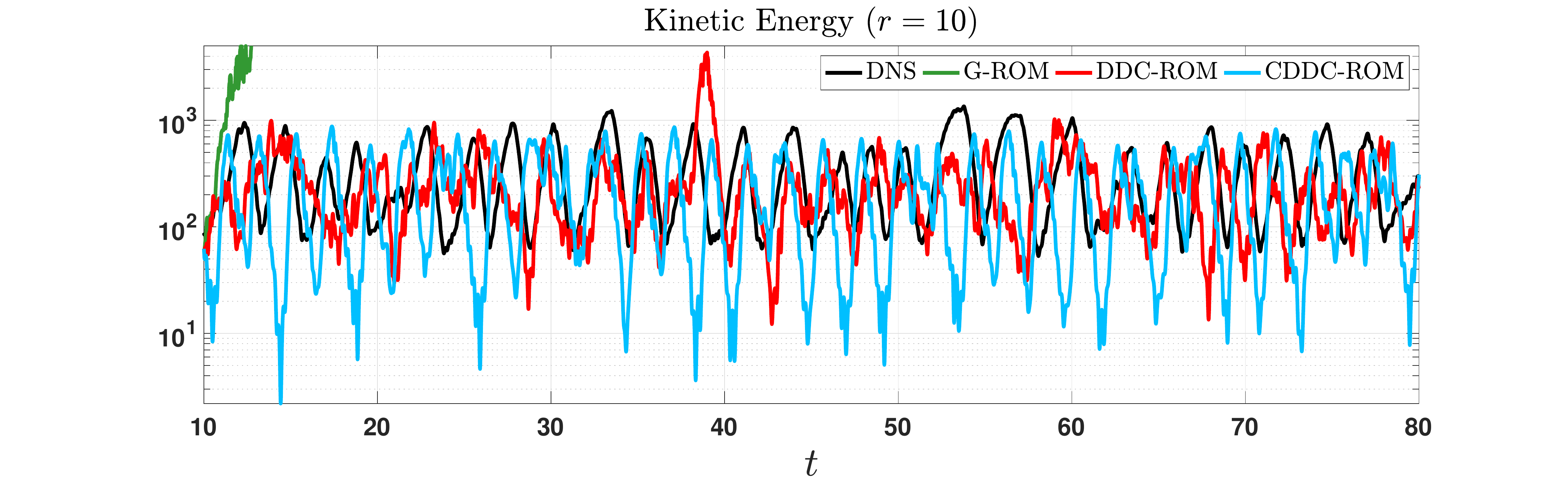}
\includegraphics[width=0.99\textwidth,height=0.25\textwidth]{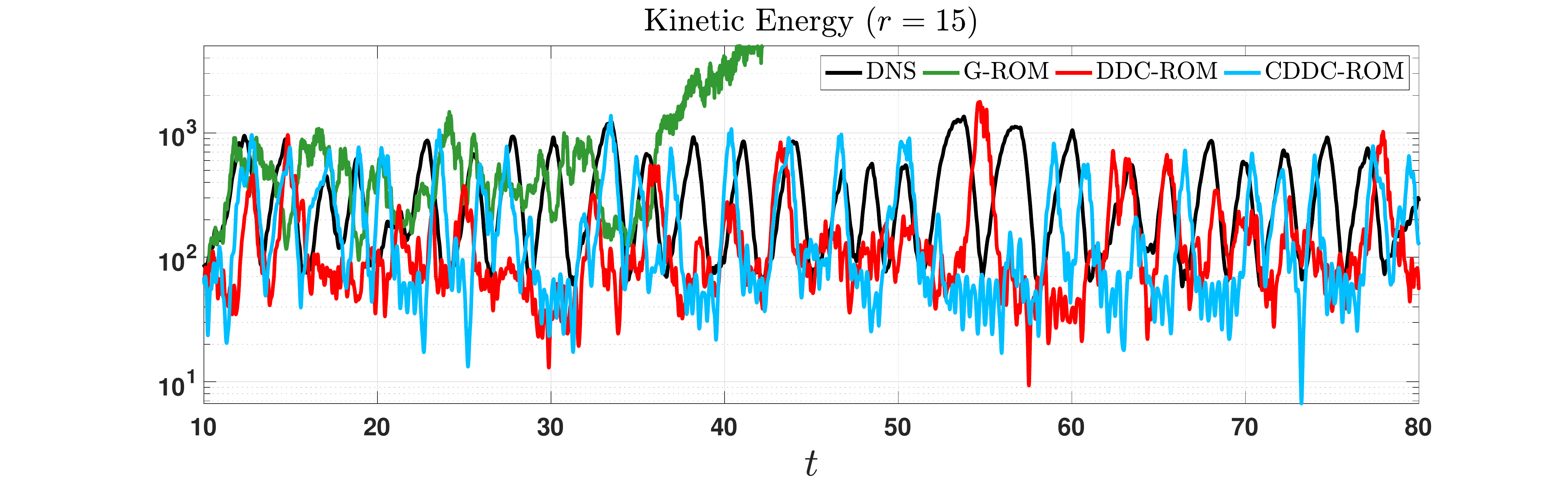} 
\includegraphics[width=0.99\textwidth,height=0.25\textwidth]{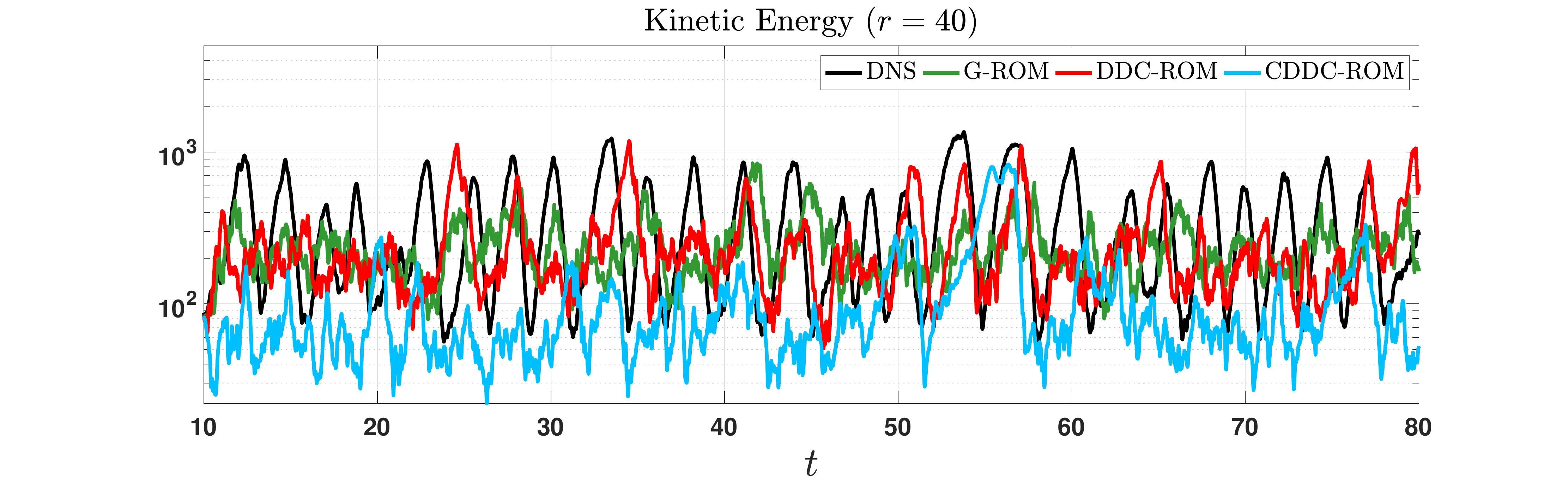}
\caption{Kinetic energy of DNS, G-ROM, DDC-ROM and CDDC-ROM with different $r$ values. All the ROMs are initialized at $t=10$ using the projected DNS data. 
}\label{fig:kinetic_energy}
\end{figure}

In Fig.~\ref{fig:kinetic_energy}, for three different $r$ values ($r=10, 15$, and $40$), we plot the time evolution of the ROM kinetic energy.
For $r=10$, the G-ROM kinetic energy takes off very quickly and stabilizes at a level around $8\times10^4$, which is roughly 200 times higher than the DNS kinetic energy on average.
In contrast, both the DDC-ROM and CDDC-ROM successfully stabilize the G-ROM, and produce kinetic energies almost of the same order of magnitude as the DNS kinetic energy (although there is some overdamping in the CDDC-ROM result due to the physical constraint $\ba^{\top} \tA \, \ba \le 0$; see \eqref{eqn:least-squares-qge-constrained}).

For $r=15$, the G-ROM kinetic energy is within good range at the beginning of the simulation, but increases to an unphysical value around $t=40$ and eventually stabilizes and oscillates around $2\times 10^4$.
In contrast, the DDC-ROM and CDDC-ROM kinetic energies are both within the good range, and between the two, the CDDC-ROM performs better in reproducing the peaks and the corresponding frequency of the peaks.

Finally, for $r=40$, the G-ROM and DDC-ROM perform similarly. The DDC-ROM kinetic energy is closer to the DNS kinetic energy over certain time windows (e.g., $[50,60]$),  whereas the CDDC-ROM kinetic energy is somewhat lower than the DNS kinetic energy.

The above results suggest that both the DDC-ROM and CDDC-ROM can successfully stabilize a severely truncated G-ROM. For the chosen criterion, the advantage of the DDC-ROM over the G-ROM is clearly visible for all ROM dimensions $r$ between $10$ and $30$. The CDDC-ROM can produce results comparable or even better than DDC-ROM for $r$ between $10$ and $20$. For even higher dimensions, the CDDC-ROM tends to overdamp the kinetic energy. This is plausible, since the physical constraint $\ba^{\top} \tA \, \ba \le 0$ in the estimation of the matrix $\tA$ for the CDDC-ROM aims to enhance the stability of the ROM, but does not also guarantee improved accuracy compared to the DDC-ROM.


\subsubsection{Relative errors for the time-averaged streamfunction}

In this section, we assess the ROM performance at a more quantitative level using the ROM time-averaged streamfunction over the aforementioned time interval, $[10, 80]$. It is known that the time-averaged streamfunction displays a four-gyre structure \cite{greatbatch2000four} even though a double-gyre wind forcing is employed; cf.~\eqref{eq:qge:forcing}. The metric that we use is the following relative error:
\begin{equation} \label{Eq_err_streamfunc}
\bigl \|\overline{\psi^{DNS}({\bf x}, \cdot)}-\overline{\psi^{ROM}({\bf x, \cdot})} \bigr \|^2_{L^2} \, \bigl/ \, \bigl\|\overline{\psi^{DNS}({\bf x}, \cdot)} \bigr\|^2_{L^2},
\end{equation}
where $\overline{(\cdot)}$ represents time average over $[10,80]$, and ${\bf x} = (x,y)$.

\begin{figure}[htp]
\centering
\includegraphics[width=0.24\textwidth,height=0.35\textwidth]{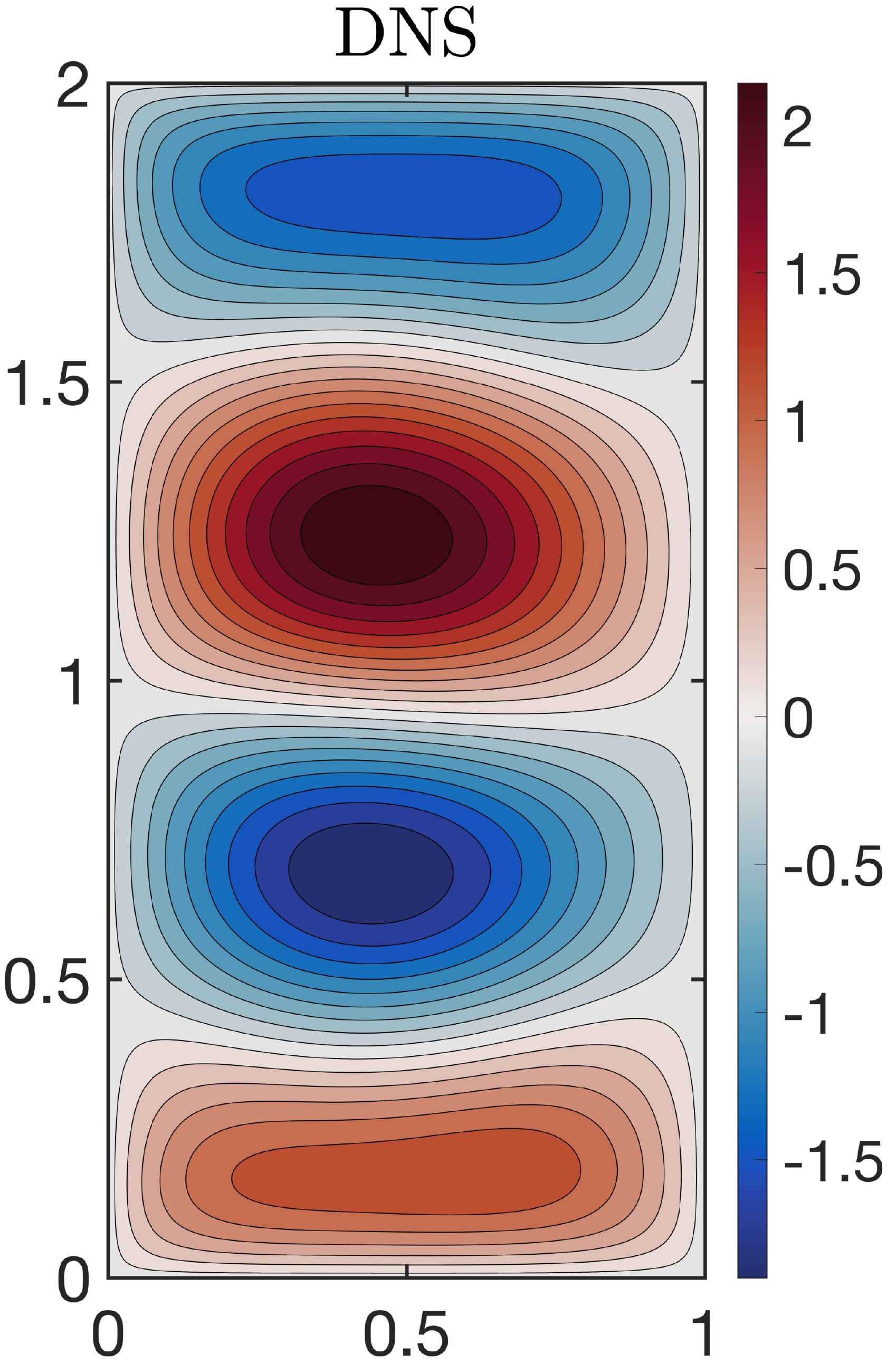} 
\includegraphics[width=0.24\textwidth,height=0.35\textwidth]{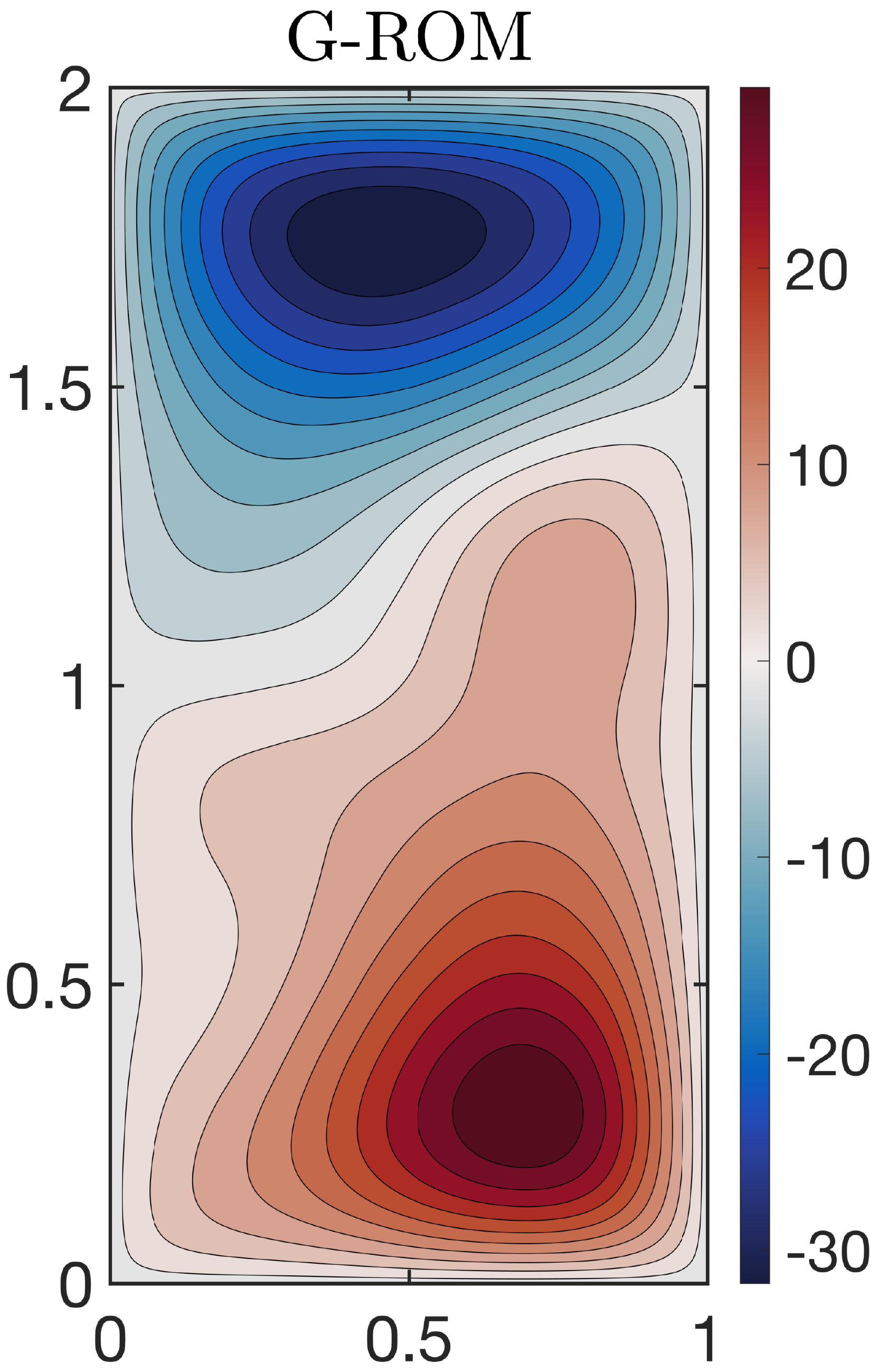} 
\includegraphics[width=0.24\textwidth,height=0.35\textwidth]{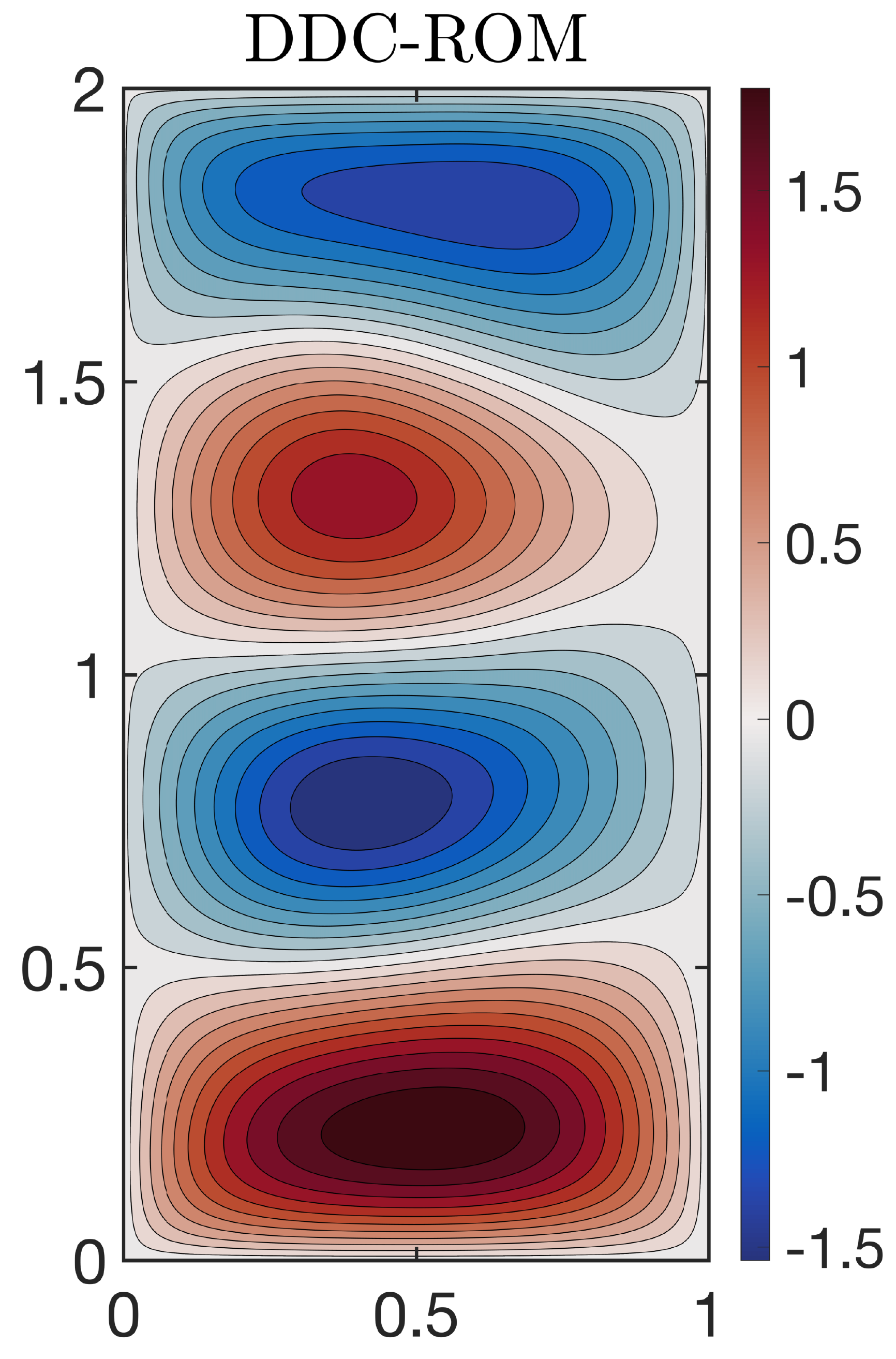}
\includegraphics[width=0.24\textwidth,height=0.35\textwidth]{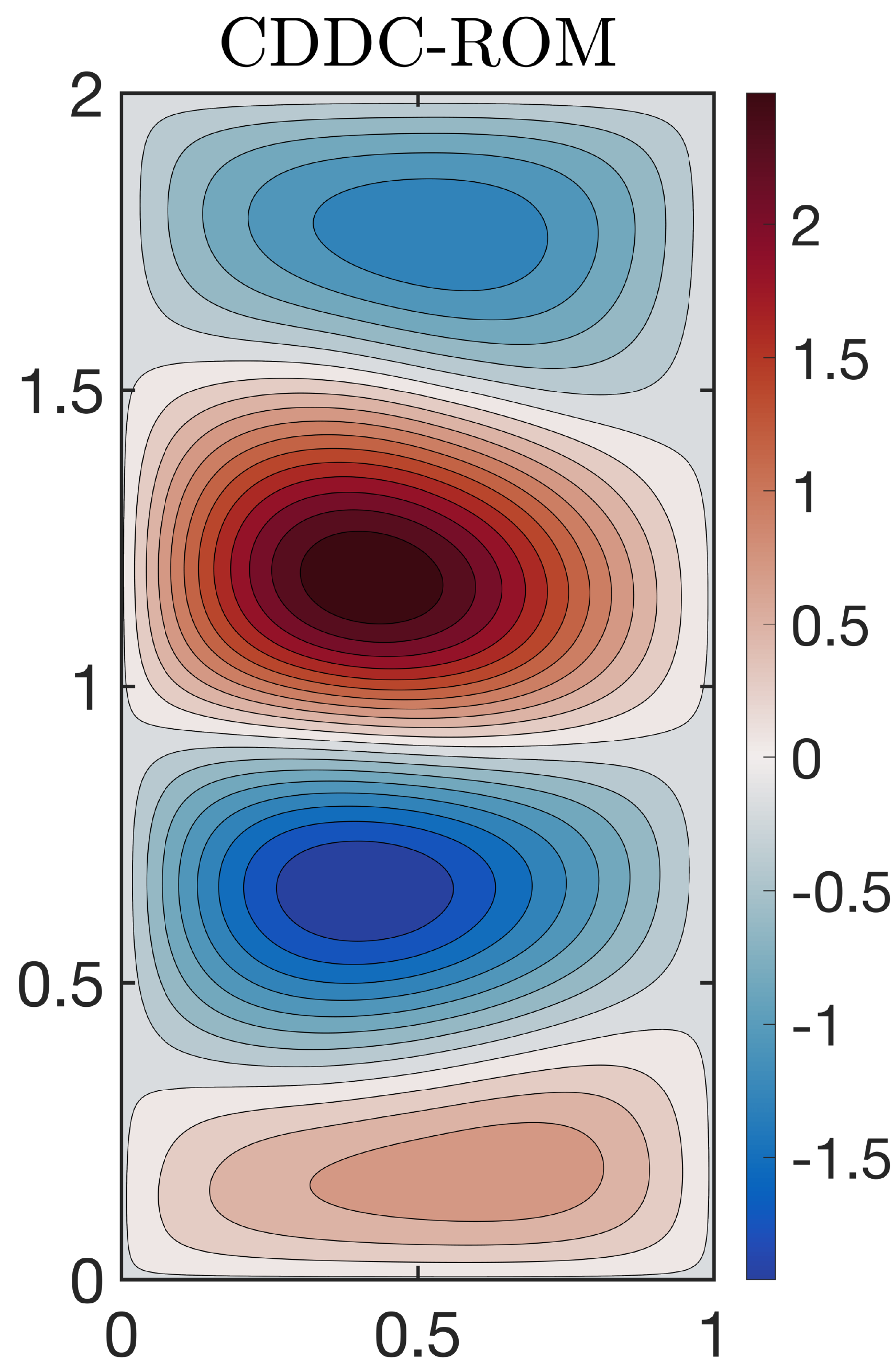}
\caption{Time-averaged streamfunction $\psi$ over the interval $[10,80]$ for DNS, $10$-dim G-ROM, $10$-dim DDC-ROM, and $10$-dim CDDC-ROM.}\label{fig:mean_fields}
\end{figure}

In Table~\ref{table:error}, we list this relative error for each of the ROMs as the ROM dimension $r$ increases. 
For small $r$ values (i.e., $5 \leq r \leq 20$), the CDDC-ROM is the most accurate.
Indeed, for $r=5$, the CDDC-ROM is the only ROM that yields a stable approximation: all other ROMs with \(r = 5\) experience exponential blowup with the given timestep.
Furthermore, for $r=10$ and $r=15$, the CDDC-ROM error is at most half of the DDC-ROM error.
Finally, for $5 \leq r \leq 50$, the G-ROM error is one to three orders of magnitude larger than the CDDC-ROM error. 
These results are also supported by the plots in Fig.~\ref{fig:mean_fields}, which display the time-average of the streamfunction $\psi$ over the time interval $[10,80]$ for DNS, G-ROM ($r=10$), DDC-ROM ($r=10$), and CDDC-ROM ($r=10$). 
These plots clearly show that the DDC-ROM and CDDC-ROM are able to capture the correct four-gyre structure, whereas the G-ROM fails drastically at this low ROM dimension.

For large $r$ values (i.e., $25 \leq r \leq 50$), the DDC-ROM results in Table~\ref{table:error} are the most accurate.
Indeed, at $r=25$ the CDDC-ROM error starts to increase, whereas the DDC-ROM error generally decreases.
The G-ROM error also continues to decrease, but it is always larger than the DDC-ROM error.

We conclude that the CDDC-ROM is the most accurate for small $r$ values and the DDC-ROM is the most accurate for large $r$ values.
These results suggest that adding physical constraints to the DDC-ROM is beneficial in the highly  truncated cases (i.e., for small $r$ values), but the benefit brought by the physical constraints diminishes as $r$ is further increased, and can even produce less accurate results than DDC-ROM due to overdamping, as pointed out in Section~\ref{sect_kinetic_energy}.
We emphasize, however, that we are using the linear ansatz to construct the DDC-ROM; further numerical investigations are needed to determine the role of physical constraints when the DDC-ROM is built with higher-order (e.g., quadratic) ansatz~\cite{mohebujjaman2019physically}. 
We also note that the G-ROM is consistently less accurate for all the $r$ values.
Finally, we note that, as $r$ increases, the errors for all the ROMs reach a plateau instead of converging to zero. 
We believe that this behavior is due to the roughness present in the vorticity basis functions, especially for the higher indices (see Fig.~\ref{fig:pod-basis}).

\begin{table}[htp]
\centering
\begin{tabular}{|c|c|c|c|c|c|c|c|}
\hline
\hline
$r$ values&G-ROM		&DDC-ROM		&CDDC-ROM \\ \hline             
$r=5$	&n/a			&n/a			 	&5.07e+00\\ \hline
$r=10$	&2.06e+02	&3.25e-01		   	&9.58e-02 \\ \hline
$r=15$	&3.05e+02	&2.83e-01		   	&1.03e-01 \\ \hline
$r=20$	&5.05e+00	&1.39e-01		   	&1.20e-01 \\ \hline
$r=25$	&1.73e+00	&1.61e-01		   	&2.96e-01 \\ \hline
$r=30$	&1.47e+00	&1.18e-01		   	&4.58e-01 \\ \hline
$r=35$	&6.83e-01		&1.34e-01		   	&7.69e-01\\ \hline
$r=40$	&4.69e-01		&9.17e-02		   	&4.92e-01\\ \hline
$r=45$	&2.81e-01		&4.16e-02		   	&6.83e+00\\ \hline
$r=50$	&3.66e-01		&8.20e-02		   	&4.16e-01\\ \hline
\hline    
\end{tabular}
\caption{The relative errors for the time-averaged streamfunction defined by \eqref{Eq_err_streamfunc}.} \label{table:error}
\end{table}


\subsubsection{Shorter training time interval}
In this subsection, we consider the situation when the ROMs are trained on a time interval that is shorter than the time interval over which the ROMs are tested. 
Specifically, we only use snapshots in the time interval $[10,t_p^\ast]$, sampled every 0.1 time units as before, to generate the ROM basis and construct the ROM operators $A, B,$ and $\tilde{A}$. This leads to a total number of $(t_p^\ast-10)/0.1+1$ snapshots. We investigate two different cases: (I) $t_p^\ast = 45$ and (II) $t_p^\ast = 35$.


In Table~\ref{table:error-predictive}, we list the relative errors associated with the ROMs for the time-averaged streamfunction defined in \eqref{Eq_err_streamfunc}.
The results show that the DDC-ROM is significantly more accurate than the G-ROM, especially for small $r$ values.
This is in line with the results in Table~\ref{table:error}.
Furrthermore, we note that, as expected, the shorter the time interval $[10 , t_p^\ast]$ is, the larger the ROM relative error.

\begin{table}[htp]
\centering
\begin{tabular}{|c|c|c|c|c|c|c|c|c|}
\hline
\hline
\multirow{2}{*}{$r$ values}  & \multicolumn{2}{c|}{Predictive Case I}  
 &\multicolumn{2}{c|}{Predictive Case II}  \\ \cline{2-5}
 &G-ROM&DDC-ROM&G-ROM&DDC-ROM   \\
\hline
$r=5$	&n/a			&n/a	&n/a			&n/a			\\ \hline
$r=10$	& 1.39e+04 &  3.33e-01  & 1.91e+02 &  6.56e-01 \\ \hline
$r=15$	&   9.58e+00  & 3.76e-01 &  1.94e+01 &  3.27e-01 \\ \hline
$r=20$	&  5.37e+00  & 1.35e-01  & 6.35e+00  & 1.80e-01\\ \hline
$r=25$	&   2.28e+00 &  9.47e-02 &  1.44e+00 &  3.05e-01\\ \hline
$r=30$	&  5.65e-01  & 1.32e-01  & 5.27e-01  & 1.96e-01\\ \hline
$r=35$	&  2.88e-01&   1.76e-01 &  2.26e-01 &  1.71e-01\\ \hline
$r=40$	&  2.07e-01 &  1.72e-01 &  2.44e-01  & 2.20e-01\\ \hline
$r=45$	&   2.86e-01&   2.15e-01  & 3.20e-01 &  2.34e-01\\ \hline
$r=50$	&   1.67e-01  & 2.09e-01  & 3.53e-01  & 2.72e-01\\ \hline
\hline    
\end{tabular}
\caption{The ROM relative errors for the time-averaged streamfunction defined in \eqref{Eq_err_streamfunc} for the two predictive test cases: the POD basis functions are generated using DNS snapshots over the time interval $[10,45]$ for Case I and over $[10,35]$ for Case II. The ROM simulations are carried out in the time interval $[10, 80]$. 
} \label{table:error-predictive}
\end{table}

\section{Conclusions}
	\label{sec:conclusions}

    We enhanced the standard Galerkin ROM (G-ROM) for the quasi-geostrophic
    equations (QGE) with an additional term derived from available data and a least squares
    optimization procedure. These ideas are based on
    previous work and are usually referred to as data-driven correction ROMs
    (DDC-ROMs) and constrained data-driven correction ROMs (CDDC-ROMs). The
    latter incorporate a negative semidefiniteness constraint into the
    optimization problem to preserve a fundamental property of the linear
    operator in the DDC-ROMs.

    The QGE are challenging equations that exhibit complex spatiotemporal behavior:
    we were able to significantly improve the G-ROM performance by adding an additional term (derived by optimization) to the
    linear component of the G-ROM. For a ROM with $10$ POD modes, the DDC-ROM
    lowered the error in the mean streamfunction (compared to the G-ROM) by a
    factor of about \(600\); similarly, the CDDC-ROM lowered the error by a
    factor of about \(2000\).

    In the future, we plan on investigating whether using a higher-order (e.g., quadratic) ansatz in the construction of the DDC-ROM and CDDC-ROM yields more accurate results than using a linear ansatz (i.e., the approach utilized in this paper).
     We also plan to study parameter sensitivity (on both the
    Reynolds and Rossby numbers) and examine the possibility of constructing a
    sequence of ROMs that work across a wide range of either value.

\section*{Acknowledgments}
This work is partially supported by NSF DMS-1821145 (CM,TI), NSF DMS-1616450 (HL), and NSF DMS-1450327 (DRW). 

\bibliographystyle{plain}
\bibliography{qge-ddc-rom}
\end{document}

%% file: notation.tex
\def\PP{{{\rm l}\kern - .15em {\rm P} }}
\def\PN2{{\PP_{N}-\PP_{N-2}}}

\newcommand{\R}{\mathbbm{R}}

\newcommand{\cO}{\mathcal{O}}

\newcommand{\bphi}{\boldsymbol{\varphi}}

\newcommand{\ba}{\boldsymbol{a}}

\newcommand{\bb}{\boldsymbol{b}}

\newcommand{\bff}{\boldsymbol{f}}

\newcommand{\bu}{\boldsymbol{u}}

\newcommand{\bur}{{\boldsymbol{u}}_r}

\newcommand{\bv}{\boldsymbol{v}}

\newcommand{\bx}{\boldsymbol{x}}
\newcommand{\bX}{\boldsymbol{X}}

\newcommand{\bXr}{{\bf X}^r}

\newcommand{\tA}{\tilde{A}}

\newcommand{\deleted}[1]{{}}
